# Forecasting Mobile Traffic with Spatiotemporal correlation using Deep Regression

Giulio Siracusano, *Senior Member, IEEE*, Aurelio La Corte

*Abstract*— The concept of mobility prediction represents one of the key enablers for an efficient management of future cellular networks, which tend to be progressively more elaborate and dense due to the aggregation of multiple technologies.

In this letter we aim to investigate the problem of cellular traffic prediction over a metropolitan area and propose a deep regression (DR) approach to model its complex spatiotemporal dynamics. DR is instrumental in capturing multi-scale and multi-domain dependences of mobile data by solving an image-to-image regression problem. A parametric relationship between input and expected output is defined and grid search is put in place to isolate and optimize performance. Experimental results confirm that the proposed method achieves a lower prediction error against state-of-the-art algorithms. We validate forecasting performance and stability by using a large public dataset of a European Provider.

## I. INTRODUCTION

THE last decades have been characterized by an exponential increase in mobile data usage, caused by the pervasive diffusion of smart devices and unprecedented diversity in mobile applications, and we are still at the beginning of the IoT revolution. This aspect has prompted the need for future generation of wireless networks to provide intelligent resource and traffic management, together with the provisioning of top quality services. The challenging requirements of zero latency and reliable gigabit experience are hindering the evolution towards the Fifth Generation (5G) cellular networks. For this reason, mobile traffic prediction is one active research field whose main aim is to predict the future locations and resource usage of mobile users to enable the reservation of network resources in future identified cells, facilitate proactive resource allocation, enhance energy efficiency[1], and finally deliver intelligent cellular networks [2]. Previous researches have investigated the dynamic characteristics of wireless traffic in order to make an accurate prediction [3], [4]. Notably, AutoRegressive Integrated Moving Average (ARIMA) has been applied on cellular traffic for prediction purpose. Undoubtedly, the pattern of cellular traffic exhibits a very complex behavior due to various factors, including user mobility, device heterogeneity, different communication mechanisms, usage pattern and user requirements. In addition, recent findings highlighted that temporal modeling [5] based exclusively on temporal correlation fails to predict correctly due to the relevant spatial dependency observed from the data. In order to achieve a finer representation of the complex pattern hidden in wireless traffic data, we need to analyze its related characteristics and evaluate their impact on prediction accuracy. Regarding to this, latest advances on deep learning methodologies have shown their ability to outperform conventional statistical models in traffic prediction [5], [6]. In [6] a prediction model based on deep belief network (DBN) has been proposed to represent the long-term dependence of cellular traffic, whereas Long Short Term Memory (LSTM) network has been described to exploit the spatial dependence, given by the contributions of multiple cells [5]. However, these approaches are unable to completely characterize the global spatial dependence of the traffic, and in particular the influence of adjacent cells, resulting hard to scale to the simultaneous prediction of a metropolitan area, which resembles thousands of cells. Lastly, in [7] it has been investigated the temporal and spatial dependence of the cellular traffic using deep learning, suggesting the possibility to predict the dynamics at a given time, by combining a limited number of previous records. To this aim, this letter proposes a method for cellular traffic prediction able to extract and exploit the inner multiscale spatiotemporal correlation by means of Deep Regression (DR). This method is based on a densely connected convolutional neural network (CNN) [8], which has been specifically re-designed for deep image-to-image regression. Whereas it is widely accepted the capability of the CNNs to reproduce the spatial dependence of input data [5], [7], [9], the temporal dependence is characterized by a composite pattern with multi-scale contributions. We investigate different ranges of spatiotemporal dependence by extending the approach proposed in [7] and evaluating the overall performance with different metrics. Experimental results shed light on different scales of dependencies of cellular traffic which contribute to the better understanding of the underlying mechanisms of evolution and variation from multiple factors which characterize its inner sparsity and diversity in time and space [4].

Authors Giulio Siracusano and Aurelio La Corte are with the Department of Electric, Electronic and Computer Engineering, University of Catania – Faculty of Engineering, I-95125, Catania, Italy (e-mail: giuliosiracusano@unict.it).

## II. DATASET ANALYSIS

### A. Multi-Source Dataset

The traffic dataset analyzed in this letter has been described in [9] and made available from a large telephony service provider in Europe, Telecom Italia. The dataset consists of time series of aggregated mobile traffic, containing total, internet, short message service (SMS) and call service, sent or received by users within a specific area over the city of Milan, Italy. The city is divided into a grid of $H \times W$ square elements called "cell" (here $H = W = 100$). The traffic is recorded during the period from 00:00 of 01-Nov-2013 to 00:00 of 01-Jan-2014 with a temporal interval of 10 minutes.

### B. Temporal characteristics

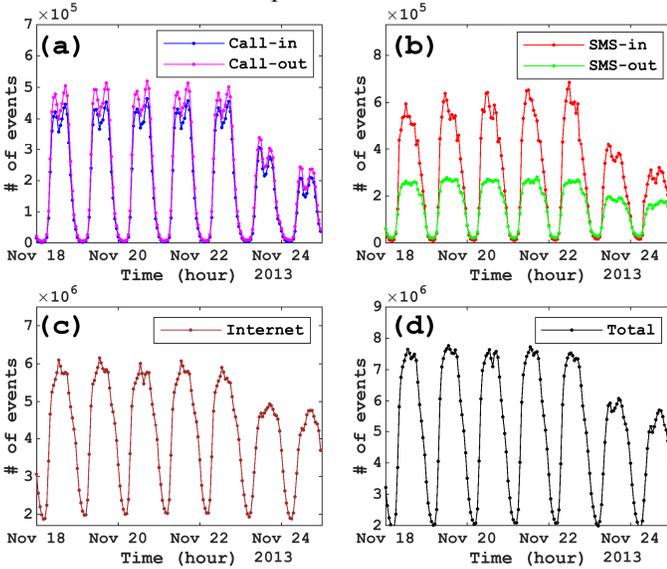

Fig. 1 – Temporal dynamics for (a) Call, (b) SMS, (c) Internet and (d) Total traffic as calculated in the period between 18 to 24 November 2013.

Based on the dataset description in [9], cellular traffic behavior is characterized by daily fluctuations of Call Detail Records (CDRs) which represent people's communication and mobility habits. Different traffic sources exhibit distinctive dynamics and characteristics. We highlight how, contrarily from data traffic in wired networks, wireless traffic possesses a higher spatiotemporal sparse characteristic [5].

Fig. 1 provides an overview of the traffic dynamics (i.e. number # of CDR events) related to different services, i.e., (a) Call, (b) SMS, (c) Internet services and the (d) total activity as computed from the entire grid. All the plot exhibits a daily periodic behavior, also when observing inbound and outbound communications such as in Call (a) and SMS (b) services, respectively. In each plot, the traffic volume drops at weekends compared with those at working days. This behavior is repeated on a weekly basis [9]. The difference between in and out traffic volumes of SMS (b) is higher than Call (a) and it is motivated by the fact that SMS are mostly generated outsize of the city area. (c) Internet service is at least an order of magnitude higher than SMS (a) and Call (b), and has also the most regular behavior which proportionally affects total traffic (d) dynamics. Taking the combined (in+out) SMS activity as an example, we investigate on its temporal correlation by calculating the average traffic volume ratio (ATVR) as a function of time. The ATVR function $\rho(\tau)$ is defined as in [7]:

$$\rho(\tau) = \frac{1}{(T-\tau) \cdot H \cdot W} \sum_{t=1+\tau}^{T} \sum_{a=1}^{H} \sum_{b=1}^{W} \frac{x_t(a,b)}{x_{t-\tau}(a,b)} \quad (1)$$

where $x_t(a,b)$ represents the traffic volume of cell $(a,b)$ at the time slot $t$ and $T$ is the number of time slots of the dataset. The value of this ratio represents how traffic at a given time is affected by the value at previous time slots. Here, values of previous time slots are more relevant than those far in time.

In Fig. 2, a plot of $\rho(\tau)$ is shown for different traffic types during 24 hours (a) and 28 days (b), respectively.

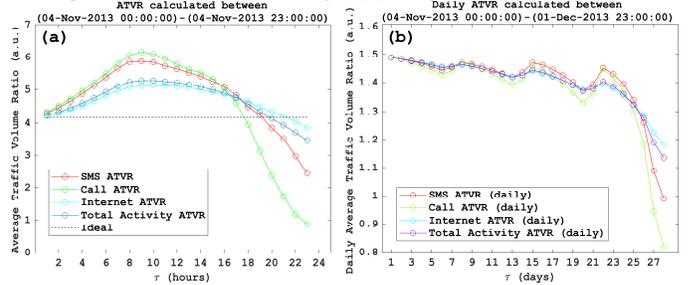

Fig. 2 – (a) Hourly ATVR calculated for SMS (red), Call (green), Internet (cyan), Total (blue) and Ideal (black dashed line) dynamics over an entire day. (b) Daily ATVR as calculated for SMS (red), Call (green), Internet (cyan) and Total traffic (blue) activity.

Fig. 2(a) reveals a cyclical behavior where the highest difference (lowest temporal correlation) is achieved after 8-9 hours and after 18-21 hours we have again that the value is close (highest temporal correlation) to the ideal plot (black dashed line) which is calculated by considering traffic of cell $(a,b)$ to be time-invariant, $x_t(a,b) \equiv x(a,b)$. When evaluated along a longer observation time (4-weeks), mobile traffics exhibit a higher temporal correlation with qualitatively similar oscillations, where, this time, the minima are localized around the week-ends because of their significantly different resource usage patterns (as shown in Fig. 1). As expected, traffic dynamics is consistent with the different human day-night activity regimes.

The study of such observed time-dependence of the traffic from previous timeslots aim us to confirm latest research [7] in terms of short-range dependence (i.e. hourly dependence from recent time fragments $t - 1, t - 2, …, t - h$), and medium-range dependence (i.e. daily dependence from same time-slot from previous days, $t - 24, t - 24 \cdot 2, …, t - 24 \cdot d$), but also let us the opportunity to highlight another longer time dependence (i.e. weekly dependence $t - 24 \cdot 7, …, t - 24 \cdot 7 \cdot w$) which we intend to take into account properly. We define hotspot the cell whose derived traffic density is larger than a given threshold. It is well known how the hotspots generally change their location during time (as a consequence of traffic variations) and it represents another study aspect of the related traffic spatial diversity. Indeed, the prediction error on the tracking of the hotspot trajectory as a function of time is considered a metric to validate the overall performance of the forecasting method [9].

*C. Spatial characteristics*: Again, considering combined SMS data, we depict in Fig. 3 a map of the spatial distribution of the traffic centered on a hotspot calculated for different periods. We measure the spatial correlation of the traffic data using a well-known metric [5], i.e., Pearson correlation coefficient $r$, between a target cell $(a,b)$ and its neighboring cells $(c,d)$:

$$r = \text{cov}(\mathbf{X}^{a,b}, \mathbf{X}^{c,d})/(s^{a,b} \cdot s^{c,d}) \quad (2)$$

where $\text{cov}(\cdot)$ represents the covariance operator and $s^{a,b}$ is the standard deviation of $\mathbf{X}^{a,b}$. For demonstration purpose, the spatial correlation of a region with 9 × 9 cells is selected to be shown. The obtained map of $r$ values is centered on the hotspot cell (59, 50) and shown in Fig. 3 for (a) 24 and (b) 168 hours, respectively.

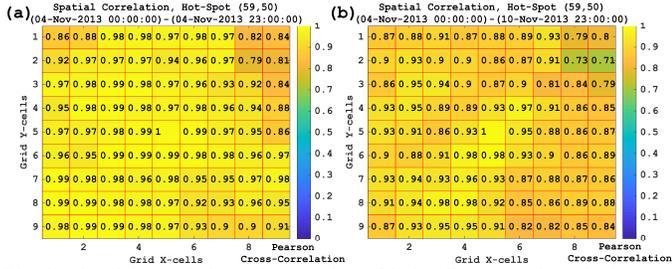

Fig. 3 – Spatial correlation $r$ for a region of 9x9 cells centered on the hotspot location (59,50) during a period of (a) 24 and (b) 168 hours, respectively.

In Fig. 3(a), for a 24-hours period, we observe a generally uniform (between 1.0 and 0.79) traffic distribution across the cells around the hotspot which tends to be globally correlated (with a slight reduction, between 1.0 and 0.71) also when considering longer observation times. For shorter time periods the higher spatial correlation among neighboring cells is a consequence of the similar services utilization that can be found in central cells, having a corresponding larger active user population if compared to others over more peripheral areas. When the observation time increases as in Fig. 3(b), the temporal dynamics reveal a still consistent correlation among the cells in the considered region and, being that Pearson function is independent from the amplitude, we obtain another relevant information about the factors which affect the traffic behavior in a longer period of time. Such findings validate the existence of an additional weekly dependence of the cellular dataset as highlighted in Fig. 2(b). Based on the above observations, it is of paramount importance to investigate on an effective method able to capture the composite spatiotemporal dependencies of the wireless traffic.

### III. MODEL AND PROCEDURE DESCRIPTION

The proposed approach consists of a sequence of three core components: (*i*) traffic preprocessing, (*ii*) training data preparation, (*iii*) Deep Regression network.

#### A. Traffic Preprocessing

The conventional method to represent a given traffic type as a function of space (grid cells) and time $t$ (time slots) is in the form of a tensor matrix $\mathbf{X}_t$, whose most intuitive interpretation is in the form of a sequence of intensity maps where each pixel corresponds to a given cell of the grid whose traffic activity can be measured for a given $t$. Traffic data are aggregated per hour to ensure prediction stability and a more memory efficient representation. A min-max normalization and standardization is applied on $\mathbf{X}_t$ to ensure zero mean and unit variance. Inverse process is applied on the predicted data during evaluation before comparing the results with the ground truth.

#### B. Training Data Preparation

For a given combination of parameters $(h, d, w)$ we can build a relationship which links the value $\mathbf{X}_t$ at time $t$, with a subset of preceding data, such that:

$$\mathbf{X}_t = f\left(\mathbf{X}_{t-i}\right)\Big|_{i=1\ldots h} \cup f\left(\mathbf{X}_{t-(168\cdot k + 24\cdot j + i)}\right)\Big|_{\substack{i=0\ldots h-1 \\ j=1\ldots d \\ k=1\ldots w}} \quad (3)$$

For example, if we set $h = d = 2$, $w = 1$, we obtain that the prediction for $\mathbf{X}_t$ is achieved by utilizing a combination of $N = h \times (d+1) \times (w+1)$ elements:

$$\mathbf{X}_t = f(\mathbf{X}_{t-1}, \mathbf{X}_{t-2}, \mathbf{X}_{t-24}, \mathbf{X}_{t-25}, \mathbf{X}_{t-48}, \mathbf{X}_{t-49}, \mathbf{X}_{t-120}, \mathbf{X}_{t-121}, \ldots \quad (4)$$
$$\ldots \mathbf{X}_{t-144}, \mathbf{X}_{t-145}, \mathbf{X}_{t-168}, \mathbf{X}_{t-169})$$

#### C. Deep Regression Network

As discussed in section II, traffic of neighboring cells may be affected by each other. For such reason, the use of densely connected CNN is motivated by its ability to hierarchically capture the spatial structural information, which is relevant for cellular traffic. In analogy with the original implementation [8], there are $L$ layers and each layer $l$ implements an operator $H_l$ which ensembles three consecutive transformations, i.e., Convolution (CV), Batch Normalization (BN) and rectified linear units (ReLU). When the CV operator is applied with a kernel of size $(k_r, k_c)$ it is able to aggregate the information of $k_r \cdot k_c$ cells leading to a progressive characterization of the global spatial dependencies of mobile service data. Our DR model is based on a variant of DenseNet-121 [8] representing a key ingredient to achieve a joint spatiotemporal characterization of traffic. Differently from recent achievements [7], we simplify processing pipeline by considering a single network which is fed with a larger combination of previous available timeslots according to Eq. (3). Here, either the bottom and top layers have been modified to handle input and output size having a section of $H \times W$ elements. All CV layers have 32 filters with size 3 × 3. The last average pooling layer is modified in 5 x 5 and it is followed by a fully-connected layer with $H \times W$ hidden units which is interposed before the new final (top) regression layer. In order to characterize spatiotemporal dependence, with the initial input tensor $\mathbf{X}^0$, at the $l^{\text{th}}$ layer, the output is the result of the recursive concatenation [8] of the features maps produced in the preceding $l - 1$ layers, such that $\mathbf{X}^l = H^l(\mathbf{X}^{l-1}) + \mathbf{X}^{l-1}$.

### IV. EXPERIMENTAL RESULTS AND ANALYSIS

In this section, we describe the testing procedures, then we evaluate and compare the results with recurrent methodologies.

#### A. Preprocessing and Parameter Settings

We extract seven full weeks from the original dataset (from 04 Nov 2013 to 29 Dec 2013), weeks 1-6 are used for training and the last one for testing purposes. In evaluation, the predicted values of the last week are unstandardized and rescaled back to the normal values and compared with the ground truth. We train the deep network using the stochastic gradient descent with momentum (SGDM) optimizer, with Momentum of 0.9, a mini-batch 128 for 50 epochs. The initial learning rate is set to 0.10, and is reduced by 20% every 10 epochs. For setting multiscale dependency parameters $h$, $d$ and *w* we use the grid search to jointly optimize our model once are defined their ranges. The root mean square error (RMSE) for the timeslot *t* is the evaluation metric defined as:

$$RMSE(t) = \sqrt{(H \cdot W)^{-1} \sum_{h=a}^{H} \sum_{w=b}^{W} \left( \gamma_{a,b}(t) - y_{a,b}(t) \right)^2} \quad (6)$$

where $\gamma_{a,b}(t)$ is the predicted value of cell $(a,b)$ whereas $y_{a,b}(t)$ is the ground truth. The overall performance is evaluated in terms of average RMSE. The results are shown in Table 1. For the considered datasets, the combination of $p = q = 3$ and $w = 1$ achieves the lowest error. This method converges in 50 epochs (results not shown). Training loss decreases abruptly during the first 10 epochs and progressively stabilizes after 30 epochs.

*Table 1 – Average RMSE for h = [2-7], d = [2-7] and w = [1-3].*

| w |   |   | 0 |   |   |   | 1 | 2 | 3 |
|---|---|---|---|---|---|---|---|---|---|
| d/h | 2 | 3 | 4 | 5 | 6 | 7 | 3 | 3 | 4 |
| 2 | 23.6 | 24.3 | 60.4 | 28.0 | 43.5 | 41.7 | 23.0 | 25.1 | 28.9 |
| 3 | 60.4 | 17.0 | 27.9 | 28.7 | 28.7 | 36.2 | **16.3** | 20.8 | 21.7 |
| 4 | 19.2 | 33.5 | 30.3 | 19.9 | 27.5 | 40.0 | 16.9 | 23.1 | 24.2 |
| 5 | 25.4 | 60.3 | 36.8 | 36.1 | 47.3 | 34.3 | 24.1 | 32.4 | 35.1 |
| 6 | 29.0 | 31.8 | 36.4 | 34.9 | 95 | 29.6 | 27.8 | 36.3 | 38.9 |
| 7 | 27.1 | 51.2 | 48.6 | 60.3 | 56.8 | 24.9 | 39.2 | 39.9 | 46.7 |

### B. Overall performance and prediction results

To validate the effectiveness of the proposed DR-based traffic prediction method, different experiments are carried out on available datasets and the results are presented in Fig. 4.

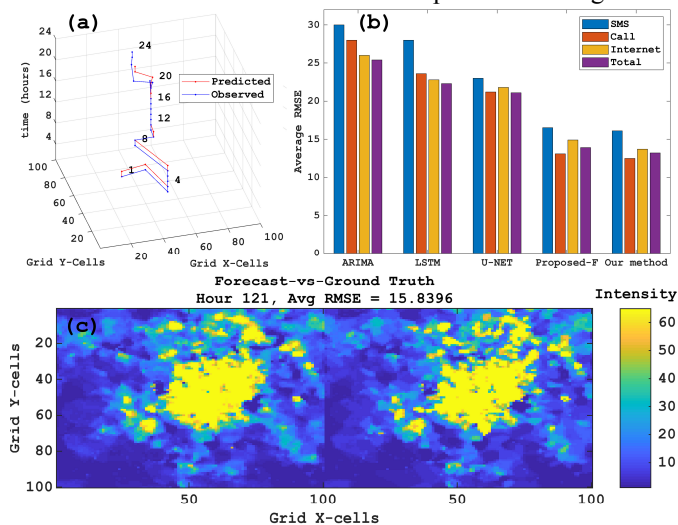

Fig. 4 – (a) Daily prediction (red line) and observed (blue line) trajectory of the hotspots from testing dataset. (b) Overall performance (average RMSE) using state-of-the-art methods for the different types of traffic. (c) Intensity images of the spatial distribution for prediction (left) and ground truth (right).

In Fig. 4(a) we show the daily prediction of hotspot (red) trajectory for SMS dataset which is evaluated against the ground truth (blue). It can be seen that the predicted trajectory well matches the ground truth over time. Only a slight difference can be seen between hours 22-23 that are associated to a minimum activity and low temporal correlation (see Fig. 2(a)) and are a consequence of weak local interactions. This evidence substantiates the forecasting stability (i.e. limited RMSE variance from different predictions) of the method.

In Fig. 4(b), the average RMSE is calculated and compared against other recurrent regression models and algorithms including ARIMA[3], LSTM[10], U-NET[11] and Proposed-F[7]. SMS dataset exhibits the highest value if compared to the other traffic types because of its least regular pattern which increases prediction error. Finally, both the predicted (left) and ground truth (right) of combined SMS traffic is provided in Fig. 4(c). The estimated spatial distribution is able to equally reproduce either amplitude and intensity of most active and less active regions preserving prediction stability over time. It is interesting to observe how also the global distribution of the traffic activity over the grid is consistent with expected data.

### V. SUMMARY AND CONCLUSIONS

This letter investigates the spatiotemporal dependence of cellular traffic and has proposed a DR approach to model such complex behavior. To accurately replicate how characteristics from different domains affect the predicted response, a parametric relationship between input and expected output has been defined and used during training and testing of the network. The obtained RMSE suggests that this method generally achieves a better prediction performance among other recurrent methods. This confirms previous accomplishments [5], [7] about the benefits of using CNNs to model an aggregate of multiple correlated cells for the study of mobile traffic. The good agreement with the prediction of hotspot trajectory highlight also that the proposed DR-based approach provide forecasting stability during the time. Our achievements prove that the proposed framework could offer a refined solution for cellular traffic characterization and prediction and significantly contribute to solve the modeling and forecasting issues.